%% file: main-public.tex
\documentclass[runningheads]{llncs}
\usepackage{graphicx}
\usepackage{hyperref}
\usepackage{amsmath}
\usepackage{amssymb}

\begin{document}
\title{Probabilistic Models of $k$-mer Frequencies\\(Extended Abstract)}
\titlerunning{Probabilistic Models of $k$-mer Frequencies}
%
\author{Askar Gafurov \and Tom\'a\v{s} Vina\v{r} \and Bro\v{n}a Brejov\'a
}
\authorrunning{A.~Gafurov et al.}
%
\institute{Faculty of Mathematics, Physics, and Informatics,
           Comenius University,\\
           Mlynsk\'a dolina, 842 48 Bratislava, Slovakia\\
           \email{\{bronislava.brejova,tomas.vinar,askar.gafurov\}@fmph.uniba.sk}}
\maketitle              
\begin{abstract}
In this article, we review existing probabilistic models for modeling abundance of fixed-length strings ($k$-mers) in DNA sequencing data.
These models capture dependence of the abundance on various phenomena, such as the size and repeat content of the genome, heterozygosity levels, and sequencing error rate.
This in turn allows to estimate these properties from $k$-mer abundance histograms observed in real data.
We also briefly discuss the issue of comparing $k$-mer abundance between related sequencing samples and meaningfully summarizing the results.


\bigskip
The final authenticated version is 
available online at \url{https://doi.org/10.1007/978-3-030-80049-9_21}

\end{abstract}

\input intro.tex
\input prelim.tex

\input spectra.tex

\input diff.tex

\input concl.tex

\bibliographystyle{splncs04}
\bibliography{main.bib}
\end{document}

%% file: intro.tex
\section{Introduction}

Rapid growth of the volume and complexity of available DNA sequencing data encourages research into efficient algorithms and data structures. A very fruitful approach is to represent individual sequences (usually sequencing reads) as sets of their subwords of a fixed length $k$ called $k$-mers. 

Many efficient methods, both exact and approximate, were developed for counting the occurrences of all $k$-mers in large sequencing datasets \cite{Manekar2018a}.
However, the focus has lately shifted to representing the sets of constituent $k$-mers without the abundance counts \cite{Chikhi2019}. This leads to reduced memory requirements, allowing representation of large collections of data sets \cite{Marchet2021}. This capability is important in the field of pangemomics, with its focus on replacing a single reference genome with a collection of individual genomes, often represented in the form of a graph built from $k$-mers occurring in these genomes \cite{Consortium2018}.

In this paper, however, we focus on $k$-mer abundance. Abundances are clearly  essential for studying transcript abundance in RNA-seq data or large-scale copy number variation \cite{Patro2014,Marchet2020}. However, usefulness of $k$-mer abundance information is not limited to these applications, but can also be used to assess fundamental properties of newly sequenced genomes.

To demonstrate this point, we review existing probabilistic models of $k$-mer abundance, which can be used to estimate genome size and other properties based on a very succinct summary of the data set---the histogram of $k$-mer abundance. In Section \ref{sec:prelim}, we define $k$-mer abundance and its spectrum. In Section \ref{sec:spectra}, we outline probabilistic models for capturing various genome and read set properties, including genome size, repeat content, heterozygosity, and sequencing error rate. Finally, in Section \ref{sec:diff},
we concentrate on comparing $k$-mer abundances in two different datasets and summarizing the results in a meaningful way. 

%% file: prelim.tex
\section{Preliminaries}
\label{sec:prelim}

A~\emph{$k$-mer} is a~string of a~length~$k$ over a~given finite alphabet~$\Sigma$; in this paper we consider the~DNA alphabet $\Sigma = \{A, C, G, T\}$.
We say that a $k$-mer $w$ \emph{matches} a sequence $S$ at position $\ell$, if $w$ is equal to the~substring of $S$ of length $k$ starting at position $\ell$.
The~number of matching positions of the~{$k$-mer $w$} in the~sequence $S$ is called the~\emph{abundance~of~$w$~in~$S$}.

Given a sequence $S$, we define an \emph{absolute $k$-mer spectrum of $S$} as the~function $h_{S,k}:\mathbb{N}\to\mathbb{N}$, where $h_{S, k}(j)$ is the~number of $k$-mers that have absolute abundance in $S$ equal to $j$.
If we normalize the~absolute $k$-mer spectrum so that the~sum of all values is one, we obtain the~\emph{relative $k$-mer spectrum of $S$}, which we denote $hr_{S, k}$.
These definitions can be easily extended from a single string $S$ to (multi)sets of strings, such as the set of chromosomes in a known genome or a set of sequencing reads.

For example, string $S=ACTACGCG$ contains dimers $CT$, $TA$, and $GC$ once, and dimers $AC$ and $CG$ twice.
Therefore, the absolute dimer spectrum has values $h_{S, 2}(1) = 3, h_{S, 2}(2) = 2$; the relative dimer spectrum has values $hr_{S, 2}(1) = 3/5, hr_{S, 2}(2) = 2/5$.
For $j > 2$, we have $h_{S, 2}(j) = hr_{S, 2}(j) = 0$.

Several variations of $k$-mers and their abundances were considered in the literature.
For example, the~quality-adjusted version of $k$-mer abundance takes into account the probabilities of sequencing errors occurring at individual positions in the sequencing reads, which are typically available in the form of base quality scores.
Each occurrence of a $k$-mer thus can be weighted by the probability that this occurrence is indeed correct \cite{Kelley2010,Comin2015}.

The notion of canonical $k$-mers helps to handle the double-stranded structure of DNA molecules. 
Both strands are usually sequenced with roughly equal probability, and therefore, it is not necessary to distinguish between a $k$-mer and its reverse complement.
The~\emph{canonical representation} of a $k$-mer $w$ is the~lexicographically smaller string among $w$ and its reverse complement.
The~\emph{absolute abundance of a canonical $k$-mer} is defined as the sum of the~absolute abundances of the~$k$-mers it represents.
While most canonical $k$-mers represent two $k$-mers, for even $k$ there are $4^{k/2}$ palindromic canonical $k$-mers that represent only one $k$-mer.
To avoid this unevenness, it is common to use only odd values of $k$ for canonical representations.

Finally, spaced $k$-mers are motifs over an extended DNA alphabet $\Sigma = \{A,C,G,T,N\}$, where $N$ stands for a ``blank'' or ``don't care'' nucleotide \cite{Morgenstern2015,Brinda2015,Rohling2019}.
We say that a spaced $k$-mer $w$ matches sequence $S$ at position $\ell$, if the~substring of $S$ of length $k$ starting at position $\ell$ agrees with $w$ at all non-blank positions.
We can consider abundances of all spaced $k$-mers that have blank symbols at predefined locations.
The main advantage of spaced $k$-mers is a smaller dependence between $k$-mer occurrences at adjacent positions of a sequence, resulting in a smaller variance of statistics used in phylogeny \cite{Morgenstern2015}.

%% file: spectra.tex
\section{Models of $k$-mer Spectra}
\label{sec:spectra}

Spectra of $k$-mer abundance represent a very compact summary of
large sequencing data sets. Assuming that genome sequencing
can be modeled as a stochastic process, the corresponding
$k$-mer spectrum will reflect important properties of the
genome, such as its size, repeat content, the level of
heterozygosity in diploid genomes, and will also depend
on the parameters of the sequencing process, such as the length of sequencing
reads, error rate, or sequencing biases.

All of these factors make modeling $k$-mer spectra an
interesting problem. In particular, given basic parameters of the
genome and the sequencing process, collectively denoted as $\theta$,
we would like to predict the corresponding $k$-mer spectrum
$hr_{\theta}$.

Such a model can then be used to interpret observed spectrum $hr$.  In
particular, our goal is to find parameters $\theta$, for which the
predicted $k$-mer spectrum $hr_{\theta}$ will be as close as
possible to the observed $k$-mer spectrum $hr$. This is typically done
by searching for $\theta^*$ minimizing a loss function, such as
cross-entropy $-\sum_i hr(i) \log
hr_{\theta^*}(i)$ or $L_2$-norm $\sum_i (hr(i) -
hr_{\theta^*}(i))^2$. Both criteria can be optimized by
general-purpose optimization algorithms supporting box constraints on parameter values,
such as L-BFGS-B \cite{Zhu1997}, and this process is typically very
efficient due to the compact data representation.

\begin{figure}
  \includegraphics[width=0.4\textwidth]{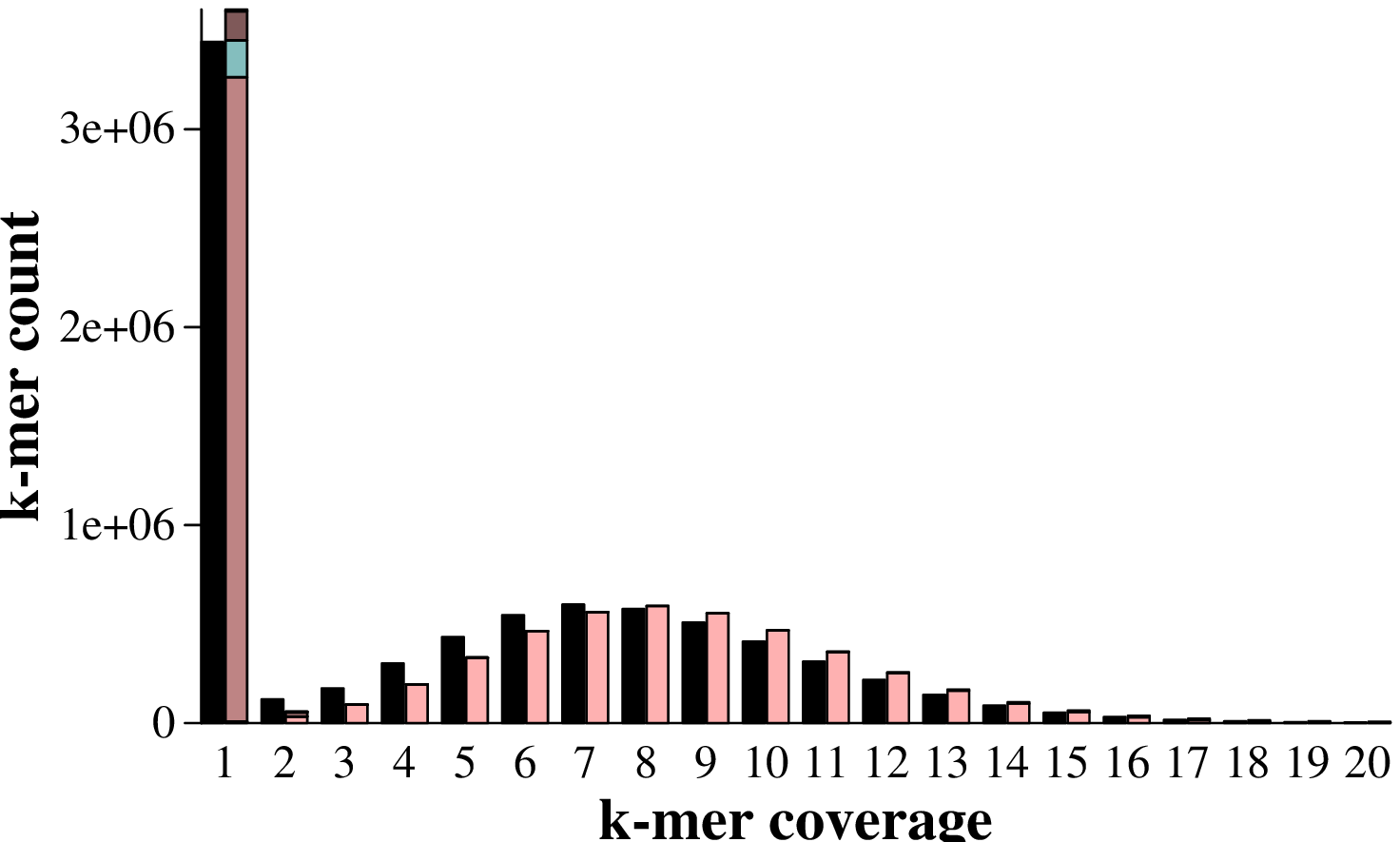}
  \includegraphics[width=0.4\textwidth]{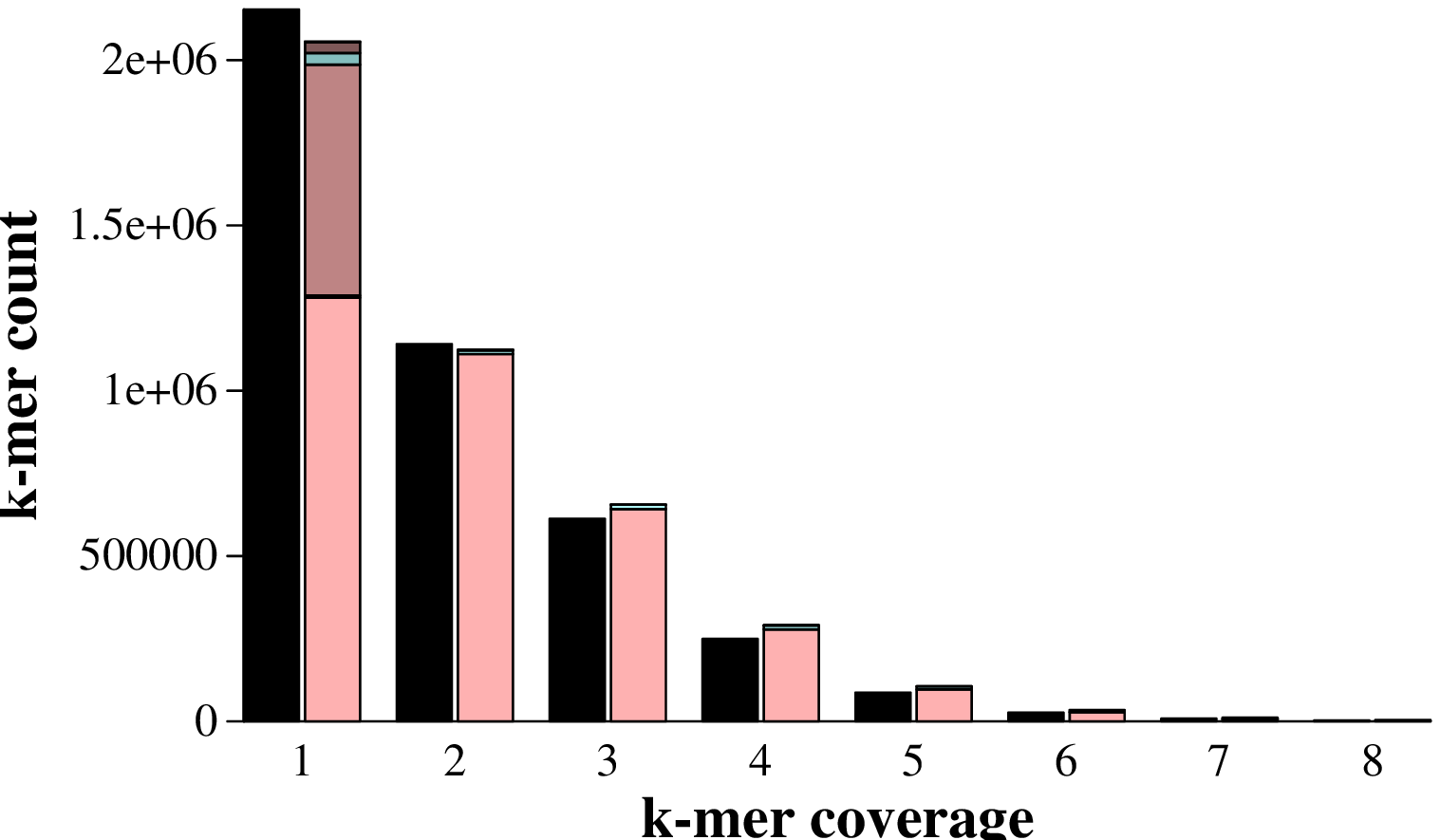}
  \raisebox{1.7cm}{\includegraphics[width=0.15\textwidth]{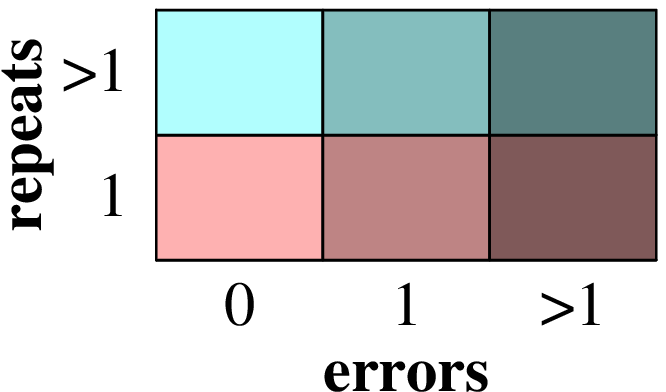}}
  \caption{Absolute 21-mer spectrum of Illumina reads for the \emph{E.~coli} genome at coverage 10 (left) and 2 (right) shown in black, and the fit of a model including sequence repeats and sequencing errors in colors. Data and model are taken from Hozza et al.~\cite{Hozza2015}.}
  \label{fig:covest}
\end{figure}

In this way, just based on the observed $k$-mer spectrum,
one can estimate key parameters of an unknown genome, such as the
genome size, without attempting a complex process of genome
assembly. Figure \ref{fig:covest}
shows 21-mer spectra for Illumina reads
produced from \emph{E.~coli} genome at $10\times$ coverage and
$2\times$ coverage.  The model used to analyze this data set contained
parameters for genome size, sequencing errors, and a simple model of
genome repeat content \cite{Hozza2015}. For high-coverage data sets,
low-abundance $k$-mers originating from sequencing errors are
clearly separable from correct $k$-mers, and thus an estimate of read
coverage can be obtained from the mode of the error-free $k$-mers
(Figure \ref{fig:covest} left).
For low-coverage data sets such a task is no longer easy 
(Figure \ref{fig:covest} right). 

In the rest of this section, we discuss models of $k$-mer
spectra, incorporating a variety of parameters representing
properties of genome sequences or the sequencing process itself.

\paragraph{A simple model.}
In the~simplest model, we assume that the~target genome is a~single circular chromosome of length $L$ with no~repeating $k$-mers, starting positions of $N$~reads are sampled uniformly independently, the~reads contain no errors, and have the~same length $r$.
Under these assumptions, the probability that a given read will cover a given $k$-mer from the genome is $p=(r-k+1)/L$, and thus the absolute abundance of each of the~$L$ $k$-mers from the genome is a random variable from the binomial distribution with parameters $N$ and $p$.
A genome with linear chromosomes behaves similarly, only $k$-mers near chromosome ends will have a smaller probability of being covered by a read. If the read length is much smaller than the chromosome lengths, this effect is negligible.

Given an observed $k$-mer spectrum (and assuming values $N$, $k$, and $r$ are known), we may therefore seek parameter $p$ of the binomial distribution that would well match the spectrum and then estimate the genome size $L$ using the value that corresponds to this value of $p$.
Note that the binomial distribution gives a non-zero probability to the event that a $k$-mer in a genome will be covered by zero reads, but such $k$-mers are not included in our observed spectra.
For low-coverage data sets, we may need to account for this observation bias by using a truncated binomial distribution.
More precisely, let $X$ be a variable from the binomial distribution, and $Y$ from the truncated distribution, where 0 cannot be observed, then $P(Y=k)=P(X=k)/P(X>0)$.

In practice, the binomial distribution can be approximated by the~Poisson distribution \cite{Simpson2014,Hozza2015} or replaced by more complex distributions to compensate for unmodelled biases.
These include the Gaussian distribution \cite{Chikhi2014,Kelley2010} and the negative binomial distribution \cite{Vurture2017,Williams2013}.
In the rest of this section, we discuss extensions of this basic model that take into account important phenomena influencing the observed spectra.

\begin{figure}[t]
  \def\RR{3.5cm}\def\SC{0.4}
  \raisebox{\RR}{(a)} \includegraphics[width=\SC\textwidth]{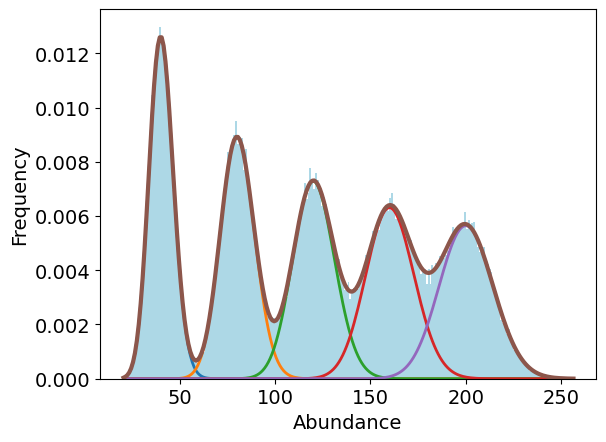}
  \raisebox{\RR}{(b)} \includegraphics[width=\SC\textwidth]{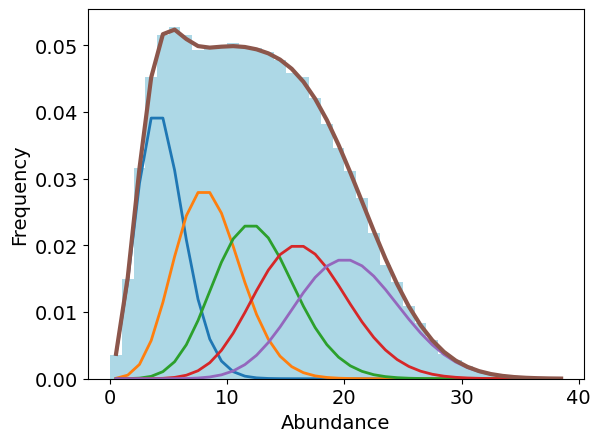}

  \caption{
    A theoretical spectrum of a genome with repeats.
    A genome with equal proportions of $21$-mers with copy numbers $1,\dots, 5$ is modeled as a mixture of binomial distributions.
    Individual distributions scaled by their weights and the mixture are shown as lines.
    The relative spectrum sampled from the mixture is shown as a light blue histogram.
    Single-copy regions of the genome have expected coverage 50 (a) and 5 (b).
  }
  
    \label{fig:spectra}
\end{figure}

\paragraph{Modeling genomic repeats.}
A significant fraction of $k$-mers in real genomes occurs in the genome more than once, due to the presence of transposons, simple tandem repeats, and segmental duplications.
Thus we have to consider the $k$-mer spectrum of the genome itself, which is usually unknown. 
The~absolute abundance of a $k$-mer in a genome is usually referred to as its \emph{copy number}.
Under the assumption of uniform sequencing, $k$-mers with higher copy numbers should appear proportionally more frequently in sequencing data.
The $k$-mer spectrum of a read set can be thus represented as a mixture of simple distributions.
Each component of the~mixture corresponds to a certain copy number and its weight is defined by the~relative genome spectrum.
Therefore, a relative read set spectrum model can be written as \[
    hr_{R}(j) = \sum_{i=1}^{\infty} hr_{S}(i) \cdot \phi(j; i, \theta),
\]
where $hr_{R}$ is the~relative read set spectrum, $hr_{S}$ is the~relative genome spectrum, and $\phi(j; i, \theta)$ is the~probability of a $k$-mer with copy-number $i$ and some shared parameters $\theta$ having exactly $j$ occurrences in the read set.
The~distribution $\phi(j;i, \theta)$ can be modeled by one of the distributions discussed for modeling genomes without repeats.
For example, when using the (truncated) Poisson distribution, parameter $\lambda_i$ for copy number $i$ will have form $\lambda_i = ic$, where $c$ is a free parameter representing coverage of single-copy $k$-mers.
At high coverage, individual components of the mixture create clearly visible peaks in the relative spectrum, but at lower coverage levels, these peaks get closer together and are more difficult to identify (Figure \ref{fig:spectra}).

It remains to model the $k$-mer spectrum of an unknown genome.
The simplest option is to let the~whole genome spectrum up to some maximum value be free parameters estimated from the data \cite{Vurture2017,Simpson2014,Williams2013}.
However, this approach has a high number of parameters, including some configurations that are not plausible.
For example, a fit of similar quality can be obtained by using copy numbers 1,2,3,\dots, or by lowering the baseline coverage and using copy numbers 2,4,6,\dots (assigning very low weights to odd copy numbers).

Another approach is to model copy numbers by a distribution, reducing the~number of free parameters.
A~popular choice is the~Zeta ($\zeta$) distribution, which has only one free parameter, governing the~shape \cite{Kelley2010,Chikhi2014}.
It is also possible to employ a~hybrid technique, where the~lowest copy-numbers are left as free parameters and the~rest is modeled as a simple parametric distribution
\cite{Hozza2015}.

\paragraph{Modeling sequencing errors.}
Sequencing errors heavily influence $k$-mer spectra of a read set by lowering the~coverage of the~true genomic $k$-mers and creating spurious $k$-mers, which typically have a low abundance (Figure \ref{fig:covest}).
One way to~handle this problem is to~discard the~lowest abundances from the~spectrum and to assume that higher abundances correspond to true $k$-mers \cite{Williams2013,Vurture2017,Ranallo-Benavidez2019}.
Another option is to~include spurious $k$-mers explicitly in the~model of the~spectrum by a~mixture model with two components corresponding to true and spurious $k$-mers, respectively.

The distribution of spurious $k$-mers can be modeled as a~parametric distribution without any interpretation and its weight can be also left as a~free parameter \cite{Kelley2010,Chikhi2014,Simpson2014}. Another option is to use a simple substitution error model \cite{Melsted2014,Hozza2015}, which assumes that every base has a probability $\varepsilon$ to be sequenced incorrectly.
Under these assumptions, the~probability of observing $k$-mer $y$ when reading $k$-mer $x$ is equal to $\varepsilon^s (1 - \varepsilon)^{k-s} 3^{-s}$, where $s$ is the~Hamming distance of $x$ and $y$. 
If $k$-mer $x$ is read $c$ times, the~expected number of times we observe $k$-mer~$y$ is equal to $\lambda_s := c\varepsilon^s (1 - \varepsilon)^{k-s} 3^{-s}$.
The~resulting distribution is then a mixture of $k+1$ distributions with means $\lambda_s$, where $s$ goes from $0$ to $k$.

\paragraph{Modeling polyploid genomes.}
Many organisms, including humans, are diploid, meaning that they have two sets of homologous chromosomes.
The~heterozygous sites in a diploid organism have two different alleles, producing two different $k$-mers instead of one.
Therefore, homozygous $k$-mers should have on average twice the~coverage of the~heterozygous ones.
This can be again represented as a mixture model, in which the~homozygous component has the~coverage parameter fixed as twice the~coverage parameter of the~heterozygous \cite{Chikhi2014,Vurture2017}.
The~weights of these components are governed by a free parameter related to the~heterozygosity of the~genome.
The situation is more complex in organisms with higher ploidy, where a single position occurs in more than two homologous chromosomes and may contain more than two alleles \cite{Ranallo-Benavidez2019}.

%% file: diff.tex
\section{Comparison of $k$-mer Frequencies Between Samples}
\label{sec:diff}

The problem of differential analysis of sequencing samples arises in many
areas: identifying differences between two individuals \cite{Shajii2016},
comparing
cancer samples with healthy tissues from the same individual \cite{Narzisi2018},
identification of differentially expressed transcripts in RNA-seq
samples \cite{Chan2019,Patro2014}, or comparing a control sample with
the sample biochemically treated to enrich or deplete particular
functional elements (such as chromatin immunoprecipitation
\cite{Zhang2008,Menzel2020} or a treatment
by enzymes depleting telomeric sequences \cite{Peska2015}).

Typically, one first aligns the reads from two or more samples to an
assembled genome or transcriptome reference and then
identifies regions (genes, transcripts, or
sequence windows) with significant differences in
read coverage between samples.
Such \emph{alignment-first} approaches work well assuming that we
have a reliable reference sequence and that we are
able to map the reads to the reference uniquely. However, in cases
where these conditions are violated, such approach may
fail. This may be because there is a structural difference between the
reference and both sequenced samples or the reference may be improperly
assembled in some regions (such as highly repetitive regions).

In this section, we concentrate on a different approach that can at
least partially overcome these limitations by comparing abundances of
individual $k$-mers between two sequencing experiments instead of
mapping the sequencing reads to the reference. Such 
approaches 
avoid potentially time-consuming alignment step, but more
importantly, they can handle repetitive regions where reads cannot be
reliably mapped and where even the assembly quality can be lower.

For example, if we sequence samples from two individuals, which share
the same expansion of a particular locus, but the number of repeats is
different from the reference sequence, the traditional reference-based
methods would still identify this region and report a false positive,
since both samples differ from the reference. On the other hand, the
alignment-free approach would correctly identify that there is no
significant difference in the $k$-mer abundances corresponding to that
locus, and thus there is no reason to highlight this repeat.

In methods based on $k$-mer abundances, we first compare the
abundances of $k$-mers in the two sets and identify $k$-mers which are
significantly underrepresented in one of the read sets \cite{Chan2019}.
Only then the
results are interpreted in the context of the reference sequence, for
example by mapping $k$-mers back to the reference
genome and identifying windows of the sequence that are significantly
enriched for the underrepresented $k$-mers. 

The window-based interpretation of the results helps us to relate
found windows to annotated genome features, such as genes, and thus to
assign them a biological meaning, which would be difficult for
individual unmapped $k$-mers. Second, by requiring multiple
underrepresented $k$-mers near each other, we filter out many false
positives, that is, individual $k$-mers that appear underrepresented
purely by chance.
Conversely, considering whole windows allows us to avoid
potential problems with local assembly errors which introduce
incorrect $k$-mers to a window. If a window is sufficiently long,
these will be compensated by the remaining correct $k$-mers. Note that this
approach is not completely alignment-free, but the alignment
to the reference is only used to interpret the results. Therefore we
call this an \emph{alignment-last} approach. Note that this window-based
method is unsuitable when searching for very
short differences, such as single nucleotide polymorphisms.

To illustrate the advantages of alignment-last methods, we
demonstrate their simple application to a simulated dataset using
chromosome IV of the yeast \emph{Saccharomyces cerevisiae} 
as a starting point.  We simulated two sequencing data sets by
selecting random substrings of length 100 and
adding substitution errors with probability $0.1\%$ at each position.
The \emph{control read set} used the reference chromosome as an
underlying string, while the second, \emph{depleted set}, was generated
from an underlying string featuring several large-scale deletions.

To simulate inaccuracies typical for draft genomes of
newly sequenced species, we used a \emph{draft assembly} produced from
simulated nanopore sequencing reads by standard methods. This draft
assembly has a single contig covering $99.8\%$ of the original chromosome IV
and has $0.5\%$ error rate.

In the baseline alignment-first method, we have mapped both read sets
to the draft assembly using Bowtie \cite{Langmead2009} and assigned
the ratio between depleted and control coverages to each base pair.
Base pairs having this ratio lower than a user-selected threshold are
marked as depleted regions. In the $k$-mer based alignment-last
method, $k$-mer abundances were counted using Jellyfish
\cite{Marcais2011}. For each $k$-mer in the genome, ratio of abundance
between the depleted and control set was computed. The draft assembly
was split into non-overlapping 100bp windows, and the score of each window
was computed as the median ratio of $k$-mers starting in this
window. Windows having the score below a user-selected threshold are
then marked as depleted regions. The accuracy of each method is summarized
by using the area under curve (AUC) statistics.

\begin{figure}
  \includegraphics[width=0.48\textwidth]{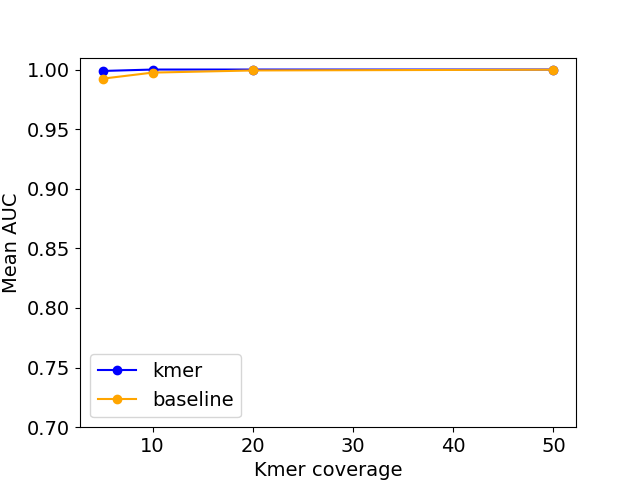}
  \includegraphics[width=0.48\textwidth]{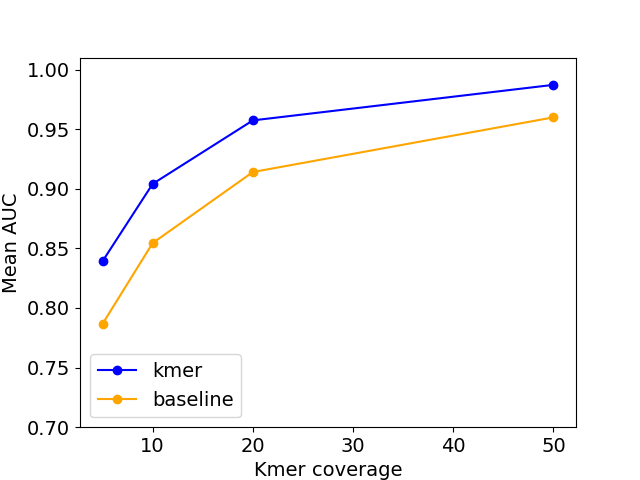}
  
  \caption{Comparison of searching for depleted regions either by the alignment-first baseline method or
    by $k$-mer abundance in windows of size 100bp on simulated data sets.
    On the left, the depleted region is a 6kb long single-copy sequence, on the right, two retrotransposons and one duplicated region of lengths 6-7kb were depleted.
    We report the AUC measure for different $k$-mer coverage levels averaged over five data sets.
  }
\label{fig:diff}  
\end{figure}

Figure \ref{fig:diff} (left) shows the results for depletion of a
6~kbp long unique sequence. For such long unique sequences both methods
can reliably identify the depleted region
even at small coverages. In Figure \ref{fig:diff} (right) we show the results
for a more complex case, where two retrotransposons and one duplicated region
of lengths between 6-7 Kbp were depleted. Here, alignment-last $k$-mer method
shows clear advantage over the baseline alignment-first method.

%% file: concl.tex
\section{Conclusion}

In this paper, we summarized techniques used in various published models of $k$-mer spectra.
Although the models cover many phenomena influencing $k$-mer abundance, some issues still remain to be explored.
One example is the influence of GC content on read coverage, which is taken into account in RNA-seq studies \cite{Patro2017}.
More complex errors models, taking into account indels and context biases would also be appropriate, particularly for third-generation sequencing data.

An important practical issue involves DNA molecules that are present in cells in high copy numbers, leading to increased read coverage in sequencing.
Examples include mitochondrial genomes in eukaryotes and plasmids in prokaryotes.
Repeat-aware models consider such molecules as repeats present in many copies and thus inflate the estimated genome size.
Pflug et al.~filter out mitochondrial reads before applying $k$-mer models for genome size estimation \cite{Pflug2020}, but perhaps a simple model of these short chromosomes could be incorporated instead.

Finally, it would be worthwhile to apply abundance models developed for $k$-mer spectra to the task of read set comparison. 

\paragraph{Acknowledgments.}

Our research was supported by grants from the Slovak Research and Development Agency APVV-18-0239, the Scientific Grant Agency VEGA 1/0463/20 to BB and VEGA 1/0458/18 to TV, the European Union's Horizon 2020 research and innovation program (PANGAIA project \#872539 and ALPACA project \#956229) and Comenius University Grant UK/278/2020 to AG.